\documentclass[aps,prd,twocolumn,superscriptaddress,showpacs]{revtex4}
\usepackage{epsfig,epsf}
\usepackage{amsmath}
\usepackage{amsthm}
\usepackage{amsfonts}
\usepackage{amssymb}
\usepackage{dsfont}
\usepackage{multirow}
\usepackage{appendix}
\usepackage{slashed}
\usepackage[active]{srcltx}
\usepackage{psfrag}

\begin{document}

\title{ Radial Excitations of the Decuplet Baryons }

\date{\today}
\author{T. M. ~Aliev}
\affiliation{Physics Department,
Middle East Technical University, 06531 Ankara, Turkey}
\author{K.~Azizi}
\affiliation{Physics Department, Do\u gu\c s University,
Ac{\i}badem-Kad{\i}k\"oy, 34722 Istanbul, Turkey}
\author{H.~Sundu}
\affiliation{ Physics Department, Kocaeli University, 41380 Izmit, Turkey}

\begin{abstract}
The ground and first excited states of the decuplet baryons are studied using the two-point QCD sum rule approach. The mass and residue of these states are computed and compared with the existing experimental data and other theoretical predictions. The results for the masses of the ground state particles as well as the excited $ \Delta $ and $ \Sigma^{*} $ states are in good consistency with experimental data. Our results on the excited $ \Xi^{*} $ and $ \Omega^{-} $ states reveal that the experimentally poorly known $ \Xi(1950) $ and $ \Omega^{-}(2250) $ can be assigned as the first excited states in $ \Xi^{*} $ and $ \Omega^{-} $ channels, respectively.   

\end{abstract}

\maketitle

\section{Introduction}

In recent years, the spectroscopy of hadrons is living its renaissance period due to the discovery of many new particles. The study of the spectroscopy and internal structure of hadrons is one of the main problems in hadron physics. Investigation of the properties of hadrons will give clear picture for understanding the dynamics of their excited states. The experimental study of the spectrum of excited baryons is one of the central elements of the physics programs of many accelerators. During last years, there have been made a remarkable progress to collect data on excited state hadrons at JLAB, MIT-Bates, LEGS, MAMI, ELSA, etc. In experiments, radial excitations of hadrons having the same quantum numbers as the ground states are discovered \cite{Agashe:2014kda}.

The analysis of the properties of excited hadrons represent a formidable task due to the fact that they can interact with many hadrons and it leads to the difficulty in their identification. For studying the properties of ground state hadrons the QCD sum rule approach \cite{Shifman:1978bx,Ioffe:1981kw} has occupied a special place and has been very successful. It is based  on fundamental QCD Lagrangian and all non-perturbative effects are parametrized in terms of quark and gluon condenstates. Now, the question is how successful this method is in the study of the  radially excited baryons, which carry the same quantum numbers as the ground states? In present letter we apply this method to study the excitations of decuplet baryons. It should be noted that the radial excitations of light-heavy mesons, the mass of first radial excitations of mesons and nucleon by using the least square method within the QCD sum rules method are analyzed in \cite{Gelhausen:2014jea} and \cite{Jiang:2015paa}, respectively. Recently, this method has been applied for estimation of the mass and residue of the first radial excitations of octet baryons in \cite{Aliev} and it has been found that the method is very predictive  for radial excitation of baryons as well.
The excited baryons have also been studied using lattice QCD in \cite{Edwards:2012fx,Burch:2006cc}. For a theoretical study on  the excited baryons  see for instance \cite{Aznauryan:2009da} and references therein .
  
This paper is organized as follows. In Section \ref{sec:QCDsumrule}, the mass sum rules for decuplet baryons including their first radial excitations are calculated. In Section \ref{sec:Num}, the numerical analysis of the obtained sum rules is presented. Last  section is devoted to  our summary  and conclusions.


\section{QCD sum rules for the mass and residue of  decuplet baryons including their radial excitations}

\label{sec:QCDsumrule}

In order to derive the two-point QCD sum rules required  for obtaining the mass and residue of the radially excited decuplet baryons, we consider the following correlation function: 
\begin{equation}
\Pi _{\mu \nu}(q)=i\int d^{4}xe^{iq\cdot x}\langle 0|\mathcal{T}\{\eta_{\mu
}^{D}(x)\bar{\eta}_{\nu }^{D}(0)\}|0\rangle ,  \label{eq:CorrF1}
\end{equation}
where $\eta_{\mu}^{D}(x)$ is the interpolating current  of
the decuplet  baryon and $ q $ is its four momentum. The interpolating current of the decuplet baryons coupled to the ground and excited states with the same quantum numbers can be written as
\begin{eqnarray}\label{Eq:Current}
\eta_{\mu}^{D}&=&A \epsilon^{abc} \Big\{(q_{1}^{aT}C\gamma_\mu q_{2}^{b})q_{3}^{c} + (q_{2}^{aT}C\gamma_\mu q_{3}^{b})q_{1}^{c} \nonumber \\
&+&  (q_{3}^{aT}C\gamma_\mu q_{1}^{b})q_{2}^{c} \Big\},
\end{eqnarray}
where $a, b, c$ are color indices, $C$ is the charge conjugation operator and $A$ is the normalization constant. The light quark fields $q_1$, $q_2$, $q_3$  and  the normalization constant  for different members of decuplet baryons are collected 
in table \ref{tab:quarkfield} (for details see for instance \cite{Ioffe:1981kw,  Azizi:2016ddw, Belyaev:1982sa,Belyaev:1982cd}). We shall also remark that the above current couples to both the spin--1/2 and spin--3/2 baryons with both the negative and positive parities (see for instance \citep{Wang:2010it,Wang:2010vn}). In the present study, we consider only the positive parity baryons. As we consider both the ground states and the first excited states, taking into account the negative parity baryons require implementing both the negative parity ground states and negative parity first excited states. This will require either additional sum rules obtained by applying derivatives  to those obtained from the two Dirac structures entering the calculations or the experimental knowledge on the ground state negative parity and first excited state positive/negative parity baryons. This brings more uncertainties that makes difficult the  reliable estimation of the  masses, especially in the case of first excited states. 
\begin{table}
\renewcommand{\arraystretch}{1.3}
\addtolength{\arraycolsep}{-0.5pt}
\small
\begin{tabular}{|c|c|c|c|c|}\hline \hline  &$A$ & $q_{1}$ & $q_{2}$ & $q_{3}$  \\\hline
$\Delta$ & $\sqrt{1/3}$ & d & d & u \\
$\Sigma^{*}$ & $\sqrt{2/3}$ & u & d & s \\
$\Xi^{*}$ & $\sqrt{1/3}$ & s & s & u\\
$\Omega^{-}$ & $1/3$ & s & s & s\\ \hline \hline \end{tabular}
\caption{The value of the normalization constant $A$ and the quark fields $q_1$, $q_2$, $q_3$ for the decuplet baryons.}
\renewcommand{\arraystretch}{1}
\addtolength{\arraycolsep}{-1.0pt}
\label{tab:quarkfield}
\end{table}

To derive the aimed QCD sum rules for the mass and residue of the considered baryons, the above correlation function has to be calculated using both the hadronic and OPE (operator product expansion) languages. By equating these two representations, one can get the QCD sum rules for the physical quantities of the considered baryons.

\subsection{Hadronic Representation}

The correlation function in the hadronic side is calculated in terms of the hadronic degrees of freedom which contains the physical parameters of the decuplet baryons. After insertion of a complete set of baryonic state with the same quantum numbers as the interpolating current, we get
\begin{eqnarray}
\Pi_{\mu\nu}^{\mathrm{Had}}(q)&=&\frac{\langle 0|\eta_{\mu }^D |D(q,s)\rangle \langle D(q,s)|\bar{\eta}_{\nu}|0\rangle}{m_{D}^{2}-q^{2}}
\nonumber \\
&+&\frac{\langle 0|\eta_{\mu }^D |D^{\prime}(q,s)\rangle \langle D^{\prime}(q,s)|\bar{\eta}_{\nu}|0\rangle}{m_{D^{\prime}}^{2}-q^{2}}
\nonumber \\
&+&\ldots,
\label{eq:phys}
\end{eqnarray}
with $m_{D}$ and $m_{D^{\prime}}$ being the mass of the ground and first excited states of the decuplet baryons, respectively. The dots indicates  the contributions coming from the higher states and continuum. Since the ground and first radial excitations of decuplet baryons have the same quantum numbers, their matrix elements between vacuum and one particle states are defined in similar manner, i.e., 
\begin{eqnarray}
\langle 0|\eta_{\mu }^D |D(q,s)\rangle &=&\lambda_{D}u_{\mu}(q,s),
\nonumber \\
\langle 0|\eta_{\mu }^D |D^{\prime}(q,s)\rangle
 &=&\lambda_{D^{\prime}}u_{\mu}^{\prime}(q,s),
\label{eq:Res}
\end{eqnarray}
where $ \lambda_{D^{(\prime)}} $ is the residue of the corresponding baryon. By 
using the summations  over spins of the Rarita-Schwinger spinor as
\begin{eqnarray}\label{Rarita}
\sum_s  u_{\mu}^{(\prime)} (q,s)  \bar{u}_{\nu}^{(\prime)} (q,s) &= &-(\!\not\!{q} + m_{D^{(\prime)}})\Big[g_{\mu\nu} -\frac{1}{3} \gamma_{\mu} \gamma_{\nu} \nonumber \\
&-& \frac{2q_{\mu}q_{\nu}}{3m^{2}_{D^{(\prime)}}} +\frac{q_{\mu}\gamma_{\nu}-q_{\nu}\gamma_{\mu}}{3m_{D^{(\prime)}}} \Big],
\end{eqnarray}
for physical part we get
\begin{eqnarray}\label{PhyssSide}
\Pi_{\mu\nu}^{\mathrm{Had}}(q)&=&\frac{\lambda_{D}^{2}}{q^{2}-m_{D}^{2}}(\!\not\!{q} + m_{D})\Big[g_{\mu\nu} -\frac{1}{3} \gamma_{\mu} \gamma_{\nu} \nonumber \\
&-& \frac{2q_{\mu}q_{\nu}}{3m^{2}_{D}} +\frac{q_{\mu}\gamma_{\nu}-q_{\nu}\gamma_{\mu}}{3m_{D}} \Big]\nonumber \\
&+&\frac{\lambda_{D^{\prime}}^{2}}{q^{2}-m_{D^{\prime}}^{2}}(\!\not\!{q} + m_{D^{\prime}})\Big[g_{\mu\nu} -\frac{1}{3} \gamma_{\mu} \gamma_{\nu} \nonumber \\
&-& \frac{2q_{\mu}q_{\nu}}{3m^{2}_{D^{\prime}}} +\frac{q_{\mu}\gamma_{\nu}-q_{\nu}\gamma_{\mu}}{3m_{D^{\prime}}} \Big]+\ldots.
\end{eqnarray}

As we previously mentioned the current $ \eta_{\mu} $ couples not only to the spin-3/2 particles but also to spin-1/2 states. Hence, the unwanted spin-1/2 contributions should be removed. To this end, we try to make the entering structures independent of each other by a special ordering of the Dirac matrices and separate the spin-1/2 contributions that can be easily remove from the correlation function. The matrix element of the  $ \eta_{\mu} $ between vacuum and spin-1/2 states can be parametrized as
\begin{eqnarray}
\langle 0 |\eta_{\mu }^D|\frac{1}{2}(q)\rangle=\left( C_1q_{\mu}+C_2\gamma_{\mu}\right) u(q),
\label{eq:spin1bol2}
\end{eqnarray}
where $ C_1 $ and $ C_2 $ are some constants and $ u(q) $ is the Dirac spinor of momentum $ q $. By imposing the condition $ \eta_{\mu }^D \gamma_{\mu}=0$, one immediately finds $ C_1 $ in terms of $ C_2 $,
\begin{eqnarray}
\langle 0 |\eta_{\mu }^D|\frac{1}{2}(q)\rangle= C_2 \left( -\frac{4}{m_{\frac{1}{2}}}q_{\mu}+\gamma_{\mu}\right) u(q),
\label{eq:spin1bol21}
\end{eqnarray}
where $ m_{\frac{1}{2}} $ is the spin-1/2 mass. It is clear from this formula that the unwanted spin-1/2 contributions are proportional to either $ q_{\mu} $ or $ \gamma_{\mu} $. After insertion of this equation into the correlation function and ordering of the corresponding Dirac matrices as $ \gamma_{\mu}\!\not\!{q}\gamma_{\nu} $ we remove the terms with the $ \gamma_{\mu} $ in the beginning, $ \gamma_{\nu} $ at the end and those proportional to $ q_{\mu} $ or $ q_{\nu} $ in order to get rid of the spin-1/2 contributions. Note that, in the results, only two structures $ \!\not\!{q}\gamma_{\mu\nu}  $ and $ g_{\mu\nu} $ contain contributions from spin-3/2 states. Hence, for hadronic part of the correlation function, we get
\begin{eqnarray}
\Pi _{\mu \nu}^{\mathrm{Had}}(q)&=&\frac{\lambda_{D}^2}{q^{2}-m_{D}^{2}} \left( \!\not\!{q}g_{\mu\nu}+m_{D}g_{\mu\nu} \right) 
\nonumber \\
&+&
\frac{\lambda_{D^{\prime}}^2}{q^{2}-m_{D^{\prime}}^{2}} \left( \!\not\!{q}g_{\mu\nu}+m_{D^{\prime}}g_{\mu\nu}\right) +\ldots,
\label{eq:CorFun1}
\end{eqnarray}
The Borel transformation with respect to $ q^{2} $, with the aim of suppressing the contributions of the higher states and continuum, leads to the final form of the hadronic  representation: 
\begin{eqnarray}
\mathcal{\widehat B}_{q^{2}}\Pi _{\mu \nu}^{\mathrm{Had}}(q)&=&\lambda_{D}^2 e^{-\frac{m_{D}^{2}}{M^{2}}} \left( \!\not\!{q}g_{\mu\nu}+m_{D}g_{\mu\nu} \right) 
\nonumber \\
&+&
\lambda_{D^{\prime}}^2 e^{-\frac{m_{D^{\prime}}^{2}}{M^{2}}} \left( \!\not\!{q}g_{\mu\nu}+m_{D^{\prime}}g_{\mu\nu}\right) +\cdots.\nonumber\\
\label{eq:CorFunBorel}
\end{eqnarray}

Here we shall remark that we perform our analysis in zero width approximation since the  resonance widths of the first radial excitations are not known yet.Taking into account these widths can bring additional uncertainties to the sum rules.

\subsection{OPE Representation}

The OPE side of the correlation function is calculated at large space-like region, where $ q^{2}\ll 0 $ in terms of  quark-gluon degrees of freedom. For this aim, we substitute the interpolating current given by Eq. (\ref{Eq:Current}) into Eq. (\ref{eq:CorrF1}), and contract the relevant quark fields. As a result, we get
\begin{widetext}
\begin{eqnarray}\label{corre1}
\Pi_{\mu\nu}^{\mathrm{OPE},\Delta}(q) &=& \frac{i}{3}\epsilon_{abc}\epsilon_{a'b'c'}\int d^4 x e^{iqx}  \langle 0 | \left\lbrace 2S^{ca'}_{d}(x)\gamma_{\nu}\widetilde{S}^{ab'}_{d}(x)\gamma_{\mu}S^{bc'}_{u}(x)-2S^{cb'}_{d}(x)\gamma_{\nu}\widetilde{S}^{aa'}_{d}(x)\gamma_{\mu}S^{bc'}_{u}(x)\right. \nonumber\\  
&+&4S^{cb'}_{d}(x)\gamma_{\nu}\widetilde{S}^{ba'}_{u}(x)\gamma_{\mu}S^{ac'}_{d}(x)+2S^{ca'}_{u}(x)\gamma_{\nu}\widetilde{S}^{ab'}_{d}(x)\gamma_{\mu}S^{bc'}_{d}(x)\nonumber\\
&-&2S^{ca'}_{u}(x)\gamma_{\nu}\widetilde{S}^{bb'}_{d}(x)\gamma_{\mu}S^{ac'}_{d}(x)-S^{cc'}_{u}(x)Tr\left[ S^{ba'}_{d}(x)\gamma_{\nu}\widetilde{S}^{ab'}_{d}(x)\gamma_{\mu}\right] \nonumber\\
&+&\left. S^{cc'}_{u}(x)Tr\left[ S^{bb'}_{d}(x)\gamma_{\nu}\widetilde{S}^{aa'}_{d}(x)\gamma_{\mu}\right] -4S^{cc'}_{d}(x)Tr\left[ S^{ba'}_{u}(x)\gamma_{\nu}\widetilde{S}^{ab'}_{d}(x)\gamma_{\mu}\right] \right\rbrace  |0 \rangle,
\end{eqnarray}
\begin{eqnarray}\label{corre2}
\Pi_{\mu\nu}^{\mathrm{OPE},\Sigma^{*}}(q) &=& -\frac{2i}{3}\epsilon_{abc}\epsilon_{a'b'c'}\int d^4 x e^{iqx} \langle 0 | \left\lbrace S^{ca'}_{d}(x)\gamma_{\nu}\widetilde{S}^{bb'}_{u}(x)\gamma_{\mu}S^{ac'}_{s}(x)\right. \nonumber\\
&+&S^{cb'}_{d}(x)\gamma_{\nu}\widetilde{S}^{aa'}_{s}(x)\gamma_{\mu}S^{bc'}_{u}(x)+S^{ca'}_{s}(x)\gamma_{\nu}\widetilde{S}^{bb'}_{d}(x)\gamma_{\mu}S^{ac'}_{u}(x)\nonumber\\
&+&S^{cb'}_{s}(x)\gamma_{\nu}\widetilde{S}^{aa'}_{u}(x)\gamma_{\mu}S^{bc'}_{d}(x)+S^{ca'}_{u}(x)\gamma_{\nu}\widetilde{S}^{bb'}_{s}(x)\gamma_{\mu}S^{ac'}_{d}(x)\nonumber\\
&+&S^{cb'}_{u}(x)\gamma_{\nu}\widetilde{S}^{aa'}_{d}(x)\gamma_{\mu}S^{bc'}_{s}(x)+S^{cc'}_{s}(x)Tr\left[  S^{ba'}_{d}(x)\gamma_{\nu}\widetilde{S}^{ab'}_{u}(x)\gamma_{\mu}\right] \nonumber\\
&+&\left. S^{cc'}_{u}(x)Tr\left[ S^{ba'}_{s}(x)\gamma_{\nu}\widetilde{S}^{ab'}_{d}(x)\gamma_{\mu}\right] +S^{cc'}_{d}(x)Tr\left[ S^{ba'}_{u}(x)\gamma_{\nu}\widetilde{S}^{ab'}_{s}(x)\gamma_{\mu}\right] \right\rbrace  |0\rangle ,
\end{eqnarray}
\begin{eqnarray}\label{corre3}
\Pi_{\mu\nu}^{\mathrm{OPE},\Xi^{*}}(q) &=& \frac{i}{3}\epsilon_{abc}\epsilon_{a'b'c'}\int d^4 x e^{iqx}\langle 0 |\left\lbrace 2S^{ca'}_{s}(x)\gamma_{\nu}\widetilde{S}^{ab'}_{s}(x)\gamma_{\mu}S^{bc'}_{u}(x)\right. \nonumber\\
&-&2S^{cb'}_{s}(x)\gamma_{\nu}\widetilde{S}^{aa'}_{s}(x)\gamma_{\mu}S^{bc'}_{u}(x)+4S^{cb'}_{s}(x)\gamma_{\nu}\widetilde{S}^{ba'}_{u}(x)\gamma_{\mu}S^{ac'}_{s}(x)\nonumber\\
&+&2S^{ca'}_{u}(x)\gamma_{\nu}\widetilde{S}^{ab'}_{s}(x)\gamma_{\mu}S^{bc'}_{s}(x)-2S^{ca'}_{u}(x)\gamma_{\nu}\widetilde{S}^{bb'}_{s}(x)\gamma_{\mu}S^{ac'}_{s}(x)\nonumber\\
&-&S^{cc'}_{u}(x)Tr\left[ S^{ba'}_{s}(x)\gamma_{\nu}\widetilde{S}^{ab'}_{s}(x)\gamma_{\mu}\right] +S^{cc'}_{u}(x)Tr\left[ S^{bb'}_{s}(x)\gamma_{\nu}\widetilde{S}^{aa'}_{s}(x)\gamma_{\mu}\right] \nonumber\\
&-&\left. 4S^{cc'}_{s}(x)Tr\left[ S^{ba'}_{u}(x)\gamma_{\nu}\widetilde{S}^{ab'}_{s}(x)\gamma_{\mu}\right] \right\rbrace  |0\rangle,
\end{eqnarray}
and
\begin{eqnarray}\label{corre4}
\Pi_{\mu\nu}^{\mathrm{OPE},\Omega^{-}}(q) &=& \epsilon_{abc}\epsilon_{a'b'c'}\int d^4 x e^{iqx} \langle 0 |\left\lbrace  S^{ca'}_{s}(x)\gamma_{\nu}\widetilde{S}^{ab'}_{s}(x)\gamma_{\mu}S^{bc'}_{s}(x)\right. \nonumber\\
&-&S^{ca'}_{s}(x)\gamma_{\nu}\widetilde{S}^{bb'}_{s}(x)\gamma_{\mu}S^{ac'}_{s}(x)-S^{cb'}_{s}(x)\gamma_{\nu}\widetilde{S}^{aa'}_{s}(x)\gamma_{\mu}S^{bc'}_{s}(x)\nonumber\\
&+&S^{cb'}_{s}(x)\gamma_{\nu}\widetilde{S}^{ba'}_{s}(x)\gamma_{\mu}S^{ac'}_{s}(x)-S^{cc'}_{s}(x)Tr\left[ S^{ba'}_{s}(x)\gamma_{\nu}\widetilde{S}^{ab'}_{s}(x)\gamma_{\mu}\right] \nonumber\\
&+&\left. S^{cc'}_{s}(x)Tr\left[ S^{bb'}_{s}(x)\gamma_{\nu}\widetilde{S}^{aa'}_{s}(x)\gamma_{\mu}\right] \right\rbrace  |0\rangle ,
\end{eqnarray}
\end{widetext}
where $ \widetilde{S}(x)=CS^{T}(x)C $  and the $S_{q}^{ab}(x)$ with $q=u, d, s  $  appearing in Eqs.~(\ref{corre1})-(\ref{corre4}) is the
light quark  propagator. The explicit expression for the light quark propagator in $ x- $space has the following form 
\begin{eqnarray}
\label{eh32v18}
S_q(x)& =& {i \rlap/x\over 2\pi^2 x^4} - {m_q\over 4 \pi^2 x^2} -
{\langle \bar q q \rangle\over 12} \left(1 - i {m_q\over 4} \rlap/x \right)\nonumber\\& -&
{x^2\over 192} m_0^2 \langle \bar q q \rangle \left( 1 -
i {m_q\over 6}\rlap/x \right) \nonumber \\
&&  - i g_s \int_0^1 du \Bigg[{\rlap/x\over 16 \pi^2 x^2} G_{\mu \nu} (ux)
\sigma_{\mu \nu} \nonumber\\& -& {i\over 4 \pi^2 x^2} u x^\mu G_{\mu \nu} (ux) \gamma^\nu
 \nonumber \\
&& 
 - i {m_q\over 32 \pi^2} G_{\mu \nu} \sigma^{\mu
 \nu} \left( \ln \left( {-x^2 \Lambda^2\over 4} \right) +
 2 \gamma_E \right) \Bigg]~,\nonumber\\
\end{eqnarray}
where $\gamma_E \simeq 0.577$ is the Euler constant and $\Lambda$ is a
scale parameter.
%

By using this propagator in Eqs.~(\ref{corre1})-(\ref{corre4}) and performing the Fourier and Borel transformations as well as applying the continuum subtraction, after  very lengthy calculations, we get
 \begin{eqnarray}
 \mathcal{\widehat{B}}_{q^{2}}\Pi _{\mu \nu }^{\mathrm{OPE}}(q)&=&\widetilde{\Pi}_{1}\, \!\not\!{q}   g_{\mu \nu} + \widetilde{\Pi}_{2} \, g_{\mu\nu}
\nonumber \\ 
& +&
  \mathrm{\cdots},
 \label{PiOPE}
 \end{eqnarray}
where the functions $ \widetilde{\Pi}_{1} $ and $ \widetilde{\Pi}_{2} $, for instance for $ \Sigma^{*} $, are obtained as
\begin{widetext}
\begin{eqnarray}
&\widetilde{\Pi}_{1}^{\Sigma^{*}}&= \frac{1}{\pi^{2}}\int_0^{s_{0}} ds e^{-\frac{s}{M^2}} \left\lbrace \frac{s^2}{5\times 2^{5}\pi^{2}}+
\frac{1}{3\times 2^{2}} \left[ \langle \bar{d} d\rangle\left(m_d-2m_u-2m_s \right)
+
\langle \bar{u} u\rangle \left(m_u-2m_d-2m_s \right) 
\right. \right. 
\nonumber \\
&+& \left.\left.
\langle \bar{s} s\rangle \left(m_s-2m_u-2m_d \right)\right] -\frac{5 \langle g_s^2 G^2\rangle}{3^2\times 2^7 \pi^2} 
+\frac{7 \langle g_s^2 G^2\rangle}{3^3\times 2^4 M^4}\left[ \langle \bar{d} d\rangle \left(m_u+m_s \right) +\langle \bar{u} u\rangle \left( m_d+m_s\right) 
\right. \right. 
\nonumber \\
&+& \left.\left.
\langle \bar{s} s\rangle \left( m_u+m_d\right)\right] \mathrm{Log}\left[\frac{s}{\Lambda^{2}} \right]
\right\rbrace  
+
\frac{m_0^{2}\langle \bar{d} d\rangle}{3^2\times 2^3 \pi^2} \left(7m_u+7m_s-5m_d \right) 
\nonumber \\
&+& 
\frac{m_0^{2}\langle \bar{u} u\rangle}{3^2\times 2^3 \pi^2} \left(7m_d+7m_s-5m_u \right)
+
\frac{m_0^{2}\langle \bar{s} s\rangle}{3^2\times 2^3 \pi^2} \left(7m_u+7m_d-5m_s \right)
  \nonumber \\
&+&\frac{4}{3^2} \left( \langle \bar{u} u\rangle\langle \bar{d} d\rangle +\langle \bar{u} u\rangle\langle \bar{s} s\rangle +\langle \bar{d} d\rangle\langle \bar{s} s\rangle\right) 
-\frac{7}{3^3 M^2} m_0^2\left( \langle \bar{u} u\rangle\langle \bar{d} d\rangle +\langle \bar{u} u\rangle\langle \bar{s} s\rangle +\langle \bar{d} d\rangle\langle \bar{s} s\rangle \right)  
\nonumber \\
&+&\frac{\langle g_s^2 G^2\rangle}{3^3\times 2^4 \pi^2 M^2} \left\lbrace \langle \bar{d} d\rangle \left[ 4(m_u+m_s)-m_d \right]  +
  \langle \bar{u} u\rangle \left[ 4(m_d+m_s)-m_u \right] +\langle \bar{s} s\rangle \left[ 4(m_u+m_d)-m_s \right]    \right\rbrace 
  \nonumber \\
&+&\frac{\langle g_s^2 G^2\rangle}{3^4\times 2^7 \pi^2 M^4}m_0^2\left\lbrace \langle \bar{d} d\rangle \left[ m_d-48(m_u+m_s)\right] 
+ 
\langle \bar{u} u\rangle \left[m_u- 48(m_d+m_s) \right]
\right. 
\nonumber \\
&+&\left.
 \langle \bar{s} s\rangle \left[m_s- 48(m_u+m_d) \right]  \right\rbrace 
 +\frac{7 \langle g_s^2 G^2\rangle}{3^3\times 2^4 \pi^2 s_0 M^2}\left\lbrace \langle \bar{d} d\rangle (m_u+m_s)+\langle \bar{u} u\rangle (m_d+m_s)
\right. 
\nonumber \\
&+&\left.
\langle \bar{s} s\rangle (m_d+m_u) \right\rbrace \left[M^2+s_0 \mathrm{Log}\left[\frac{s}{\Lambda^{2}} \right]\right] e^{-\frac{s_0}{M^2}},
\label{Coefqslashgmunu}
\end{eqnarray}
and
\begin{eqnarray}
&\widetilde{\Pi}_{2}^{\Sigma^{*}}&=\frac{1}{\pi^{2}}\int_0^{s_{0}} ds e^{-\frac{s}{M^2}} \left\lbrace \frac{s^2 \left( m_u+m_d+m_s\right) }{2^{6} \pi^{2}}
- \frac{\left(\langle \bar{u} u\rangle + \langle \bar{d} d\rangle + \langle \bar{s} s\rangle \right) s}{3^2}-\frac{m_{0}^{2}\left(\langle \bar{u} u\rangle + \langle \bar{d} d\rangle + \langle \bar{s} s\rangle \right)}{3^2\times 2}
\right. 
\nonumber \\
&+&\left. \frac{\langle g_{s}^{2}G^{2}\rangle \left(m_u+m_d+m_s \right) }{3^2\times 2^7 \pi^2}\left[ \left(8\gamma_{E}-3 \right)-8 \mathrm{Log}\left[\frac{s}{\Lambda^{2}} \right]  \right]  
+
\frac{\langle g_{s}^{2}G^{2}\rangle^{2} \left(m_u+m_d+m_s \right) }{3^3\times 2^9 \pi^2 M^4}\mathrm{Log}\left[\frac{s}{\Lambda^{2}}\right]
 \right\rbrace 
 \nonumber \\
&+& \frac{2}{3} \left[ \langle \bar{d} d\rangle\langle \bar{s}s\rangle m_u+\langle \bar{d} d\rangle\langle \bar{u}u\rangle m_s+\langle \bar{d} d\rangle\langle \bar{u}u\rangle m_d \right] 
-\frac{m_0^2}{3^3\times 2^2 M^2}\left[ \langle \bar{d} d\rangle\langle \bar{s}s\rangle \left( 4m_u+9m_d+9m_s\right)
\right. 
\nonumber \\
&+&\left.
 \langle \bar{d} d\rangle\langle \bar{u}u\rangle \left( 9m_u+9m_d+4m_s\right)
+
 \langle \bar{d} d\rangle\langle \bar{u}u\rangle \left( 9m_u+4m_d+9m_s\right) \right] 
 \nonumber \\
&-& \frac{\langle g_s^2G^2\rangle}{3^2\times 2^4 \pi^{4}}M^2 \left( m_u+m_d+m_s\right) \gamma_{E} \left(1-e^{-\frac{s_0}{M^2}} \right)  
\nonumber \\
&+& \frac{\langle g_s^2G^2\rangle^{2}\left(m_u+m_d+m_s \right) }{3^4\times 2^9 \pi^{4} M^2 s_0} \left(5s_0+3M^2 e^{-\frac{s_0}{M^2}}
+
3s_0 \mathrm{Log}\left[ \frac{s_0}{\Lambda^{2}}\right]e^{-\frac{s_0}{M^2} } \right) 
\nonumber \\
&+& \frac{\langle g_s^2G^2\rangle }{3^3\times 2^3 \pi^{2}}\left(\langle \bar{u} u\rangle +\langle \bar{d} d\rangle +\langle \bar{s} s\rangle \right) 
+\frac{\langle g_s^2G^2\rangle }{3^2\times 2 M^4}\left( m_u \langle \bar{d} d\rangle \langle \bar{s} s\rangle + m_s \langle \bar{d} d\rangle \langle \bar{u} u\rangle + m_d \langle \bar{u} u\rangle \langle \bar{s} s\rangle \right) 
\nonumber \\
&-& \frac{\langle g_s^2G^2\rangle }{3^3\times 2^6 \pi^2 M^2}\left( m_0^2 \langle \bar{d} d\rangle +  m_0^2 \langle \bar{u} u\rangle +  m_0^2 \langle \bar{s} s\rangle\right) 
\nonumber \\
&+& \frac{\langle g_s^2G^2\rangle}{3^{2}\times 2 \pi^{2} M^6} \left( m_u m_0^2 \langle \bar{d} d\rangle \langle \bar{s} s\rangle +m_d m_0^2 \langle \bar{u} u\rangle \langle \bar{s} s\rangle 
+
m_s m_0^2 \langle \bar{u} u\rangle \langle \bar{d} d\rangle\right).
\label{Coefgmunu}
\end{eqnarray}
\end{widetext}
 The results for the $ \Delta $, $ \Xi^{*} $ and $ \Omega^{-} $ baryons can be obtained from  Eqs. (\ref{Coefqslashgmunu}) and (\ref{Coefgmunu}) with the help of the following replacements:
\begin{eqnarray}
\widetilde{\Pi}_{1,2}^{\Delta}&=&\widetilde{\Pi}_{1,2}^{\Sigma^{*}}\left(u\rightarrow d, s \rightarrow u \right) ,
\nonumber \\
\widetilde{\Pi}_{1,2}^{\Xi^{*}}&=&\widetilde{\Pi}_{1,2}^{\Sigma^{*}}\left(u\rightarrow s, d \rightarrow s, s \rightarrow u \right) ,
\nonumber \\
\widetilde{\Pi}_{1,2}^{\Omega^{-}}&=&\widetilde{\Pi}_{1,2}^{\Sigma^{*}}\left(u\rightarrow s, d \rightarrow s \right). 
\label{replacement}
\end{eqnarray}

Having calculated both the hadronic and OPE sides of the correlation function, we match the coefficients of the structures $ \!\not\!{q}   g_{\mu \nu} $  and $ g_{\mu \nu} $ from these two sides and obtain the following sum rules that will be used to extract the masses and residues of the ground and first excited states: 
\begin{eqnarray}
\lambda_{D}^2 e^{-\frac{m_{D}^{2}}{M^{2}}}+\lambda_{D^{\prime}}^2 e^{-\frac{m_{D^{\prime}}^{2}}{M^{2}}}&=&\widetilde{\Pi}_{1}^{D},
\nonumber \\
m_D \lambda_{D}^2 e^{-\frac{m_{D}^{2}}{M^{2}}}+m_{D^{\prime}} \lambda_{D^{\prime}}^2 e^{-\frac{m_{D^{\prime}}^{2}}{M^{2}}}&=&\widetilde{\Pi}_{2}^{D}.
\label{Eq:sumrule}
\end{eqnarray}


\section{Numerical results}

\label{sec:Num}

As is seen from Eq. (\ref{Eq:sumrule}) in order to obtain the numerical values of the mass and residue of the radial excitations of the decuplet baryons, we need to have the values of the mass and residue of the ground states. Because of this, firstly, we calculate the mass and residue of the  ground state particles by choosing an appropriate threshold $ s_0 $  according to the standard prescriptions. The working region of the threshold parameter is found requiring that the sum rules show a good convergence of the OPE and lead to a maximum possible pole contribution. Besides, the physical quantities are demanded to demonstrate relatively weak dependencies on this parameter in its working interval.
 The sum rules contain numerous parameters, i.e. quark, gluon and mixed
condensates and mass of the light quarks, values of which are
shown in table \ref{tab:Param}. 

\begin{table}[tbp]
\begin{tabular}{|c|c|}
\hline\hline
Parameters & Values \\ \hline\hline
$m_{u}$ & $2.2^{+0.6}_{-0.4}~\mathrm{MeV}$\cite{PDG} \\
$m_{d}$ & $4.7^{+0.5}_{-0.4}~\mathrm{MeV}$ \cite{PDG}\\
$m_{s}$ & $96^{+8}_{-4}~\mathrm{MeV}$ \cite{PDG}\\
$\langle \bar{q}q \rangle $ & $(-0.24\pm 0.01)^3$ $\mathrm{GeV}^3$ \cite{Belyaev:1982sa}  \\
$\langle \bar{s}s \rangle $ & $0.8\langle \bar{q}q \rangle$ \cite{Belyaev:1982sa} \\
$m_{0}^2 $ & $(0.8\pm0.1)$ $\mathrm{GeV}^2$ \cite{Belyaev:1982sa}\\
$\langle g_s^2 G^2 \rangle $ & $4\pi^2 (0.012\pm0.004)$ $~\mathrm{GeV}
^4 $\cite{Belyaev:1982cd}\\
$ \Lambda $ & $ (0.5\pm0.1) $ $\mathrm{GeV} $ \cite{Chetyrkin:2007vm} \\
\hline\hline
\end{tabular}%
\caption{Some input parameters.}
\label{tab:Param}
\end{table}
\begin{table}[tbp]
\begin{tabular}{|c|c|c|}
\hline\hline
Baryon & $ M^2 \,\mathrm{(GeV^2)}$ & $ s_0 \,\mathrm{(GeV^2)} $ \\ \hline\hline
$\Delta$ & $1.5\leq M^2 \leq3.0$&$1.7^2\leq s_0 \leq 1.9^2$ \\ \hline
$\Sigma^{*}$ & $1.6\leq M^2 \leq3.5$&$1.8^2\leq s_0 \leq 2.0^2$ \\ \hline
$\Xi^{*}$ & $1.9\leq M^2 \leq4.0$&$2.0^2\leq s_0 \leq 2.2^2$ \\ \hline
$\Omega^{-}$ & $2.0\leq M^2 \leq 5.0$&$2.1^2\leq s_0 \leq 2.3^2$ \\ \hline
\hline\hline
\end{tabular}%
\caption{The working regions of $ M^2 $ and $ s_0 $ for  the ground state  $ \Delta $, $ \Sigma^{*} $, $ \Xi^{*} $ and $ \Omega^{-}  $ baryons.}
\label{tab:GSM2s0}
\end{table}

In addition to the input parameters, the auxiliary parameter $ M^2 $ should also be fixed. We find the working region of  $ M^2 $ such that the physical quantities  weakly depend on   it as much as possible.  This is achieved demanding that not only the contributions of the higher states and continuum should be small compared to the ground  and first excited states contributions, but also the higher dimensional operators should have small contributions and the series of sum rules should converge.   

Using the above procedures, the working regions of  $ M^2 $ for different channels are obtained. These intervals together with the working regions of $ s_0 $ for the ground states  are given in table \ref{tab:GSM2s0}. After standard analysis of the sum rules, we extract the values of the mass and residue of the ground state decuplet baryons as presented in table \ref{tab:GSNumValues}. Note that the presented values are obtained taking the average of the corresponding values obtained via two different sum rules presented in Eq. (20). The two some rules' predictions differ   with amount of maximally $ 5\% $ with each other that we have included in the errors. We also compare our results with the existing experimental data and other theoretical predicting in this table. From this table, we see that our predictions on the ground state mass of the decuplet baryons are in good consistencies with the average experimental data presented in PDG \cite{PDG}. In the case of the residues, our predictions for the residue of  ground states  are overall comparable with those obtained in \cite{Azizi:2016ddw,Lee:1997ix}.

\begin{widetext}

\begin{table}[tbp]
\begin{tabular}{|c|c|c|c|c|}
\hline\hline
 Mass & $ m_{\Delta} \,\mathrm{(MeV)}$ &$ m_{\Sigma^{*}} \,\mathrm{(MeV)}$ & $ m_{\Xi^{*}} \,\mathrm{(MeV)}$ &$ m_{\Omega^{-}} \,\mathrm{(MeV)}$ 
  \\ \hline\hline
Present study &$1226\pm 124  $  & $ 1389\pm142 $&$1577\pm163  $ & $1657\pm172  $ \\ \hline
Experiment \cite{PDG} & $ 1209 - 1211 $ & $  1382.80\pm0.35$ & $ 1531.80\pm0.32 $ & $1672.45\pm0.29  $ \\ 
\hline\hline\hline
Residue  & $ \lambda_{\Delta} \,\mathrm{(GeV^3)} $&$ \lambda_{\Sigma^{*}} \,\mathrm{(GeV^3)} $ & $ \lambda_{\Xi^{*}} \,\mathrm{(GeV^3)} $ &$ \lambda_{\Omega^{-}} \,\mathrm{(GeV^3)} $
  \\ \hline\hline
Present study & $ 0.029\pm0.008 $ & $ 0.036\pm0.010 $ &$ 0.045\pm0.013 $ & $ 0.049\pm0.015 $\\ \hline
\cite{Azizi:2016ddw} & $ 0.038\pm0.010$& $ 0.043\pm0.012 $&$ 0.053\pm0.014 $& $ 0.068\pm0.019$
\\ \hline
 \cite{Lee:1997ix}(average central values) &$ 0.044 $& $ 0.051 $& $ 0.062 $&$0.073  $
\\ \hline\hline
\end{tabular}%
\caption{The numerical values of masses and residues of  the ground state  $ \Delta $, $ \Sigma^{*} $, $ \Xi^{*} $ and $ \Omega^{-} $  baryons. In \cite{Lee:1997ix} the numerical values of $ \widetilde{\lambda}^{2}=4\pi^{2}\lambda^{2} $ are given in chiral-odd and chiral-even sum rules approaches. To compare with our results, we calculate the average $ \lambda $ from those results and do not show the uncertainties of the results. } 
\label{tab:GSNumValues}
\end{table}

\end{widetext}
\begin{table}[tbp]
\begin{tabular}{|c|c|}
\hline\hline
Baryon & $ s'_0 \,\mathrm{(GeV^2)}$ \\ \hline\hline
$\Delta'$ &$2.2^2\leq s^{\prime}_0 \leq 2.4^2$ \\ \hline
$\Sigma^{*'}$&$2.5^2\leq s^{\prime}_0 \leq 2.7^2$ \\ \hline
$\Xi^{*'}$ &$2.8^2\leq s^{\prime}_0 \leq 3.0^2$ \\ \hline
$\Omega^{-'} $ &$3.1^2\leq s^{\prime}_0 \leq 3.3^2$ \\ \hline
\hline
\end{tabular}%
\caption{The working regions of  $ s^{\prime}_0 $ for  the radially excited $ \Delta' $, $ \Sigma^{*'} $, $ \Xi^{*'} $ and $ \Omega^{-'} $  baryons.}
\label{tab:REM2s0}
\end{table}
\begin{widetext}

\begin{table}[tbp]
\begin{tabular}{|c|c|c|c|c|}
\hline\hline
 Mass & $ m_{\Delta^{\prime}} \,\mathrm{(MeV)}$ &$ m_{\Sigma^{*\prime}} \,\mathrm{(MeV)}$ & $ m_{\Xi^{*\prime}} \,\mathrm{(MeV)}$ &$ m_{\Omega^{-\prime}} \,\mathrm{(MeV)}$ 
  \\ \hline\hline
Present study &$1483\pm 133  $  & $ 1719\pm179 $&$1965\pm178  $ & $2176\pm219  $ 
\\ \hline
Experiment \cite{PDG} & $1460\,-\, 1560$ & $ 1727\pm27 $  & - & -
 \\ 
\hline\hline\hline
Residue  & $ \lambda_{\Delta^{\prime}} \,\mathrm{(GeV^3)} $&$ \lambda_{\Sigma^{*\prime}} \,\mathrm{(GeV^3)} $ & $ \lambda_{\Xi^{*\prime}} \,\mathrm{(GeV^3)} $ &$ \lambda_{\Omega^{-\prime}} \,\mathrm{(GeV^3)} $
  \\ \hline\hline
Present study & $ 0.057\pm0.016 $ & $ 0.076\pm0.022 $ &$ 0.103\pm0.030 $ & $ 0.129\pm0.039 $
\\ \hline\hline\hline
\end{tabular}%
\caption{The numerical values of masses and residues of the radially excited  $ \Delta' $, $ \Sigma^{*'} $, $ \Xi^{*'} $ and $ \Omega^{-'} $  baryons.}
\label{tab:RDNumValues}
\end{table}

\end{widetext}

At this stage, we proceed to find the values of the masses and residues of the radially excited baryons considering the values of the masses and residues of the corresponding ground state baryons as inputs. For this aim the sum rules in Eq. (\ref{Eq:sumrule}) are used. 
When taking into account the contributions of the radial excitations,  the continuum threshold  should be changed compared to the previous case. Using the standard criteria, the continuum threshold $ (s^{\prime}_{0}) $   including the  first excited states is found as presented in table \ref{tab:REM2s0}. 
Our analyses show that by the above intervals for $ M^{2} $,  $ s_0 $ and $ s^{\prime}_{0} $, the OPE nicely converges and the pole ground and first excited states constitute  the main part of the total result and the effects of  higher states and continuum are relatively small. Taking the average of the values in the channels under consideration, the pole and first excited state contribution is found to be $ 59\% $ of the total contribution. 
The mass and residue of the excited $ D^{\prime} $ baryons versus $ M^{2} $ at different values of new continuum threshold are shown in Figs.\ref{REDelta}-\ref{REOmega}. Extracted from our analysis, we depict the numerical values of the masses and residues of $ D^{\prime} $ baryons in table \ref{tab:RDNumValues}.

\begin{widetext}

\begin{figure}[h!]
\begin{center}
\includegraphics[totalheight=5cm,width=7cm]{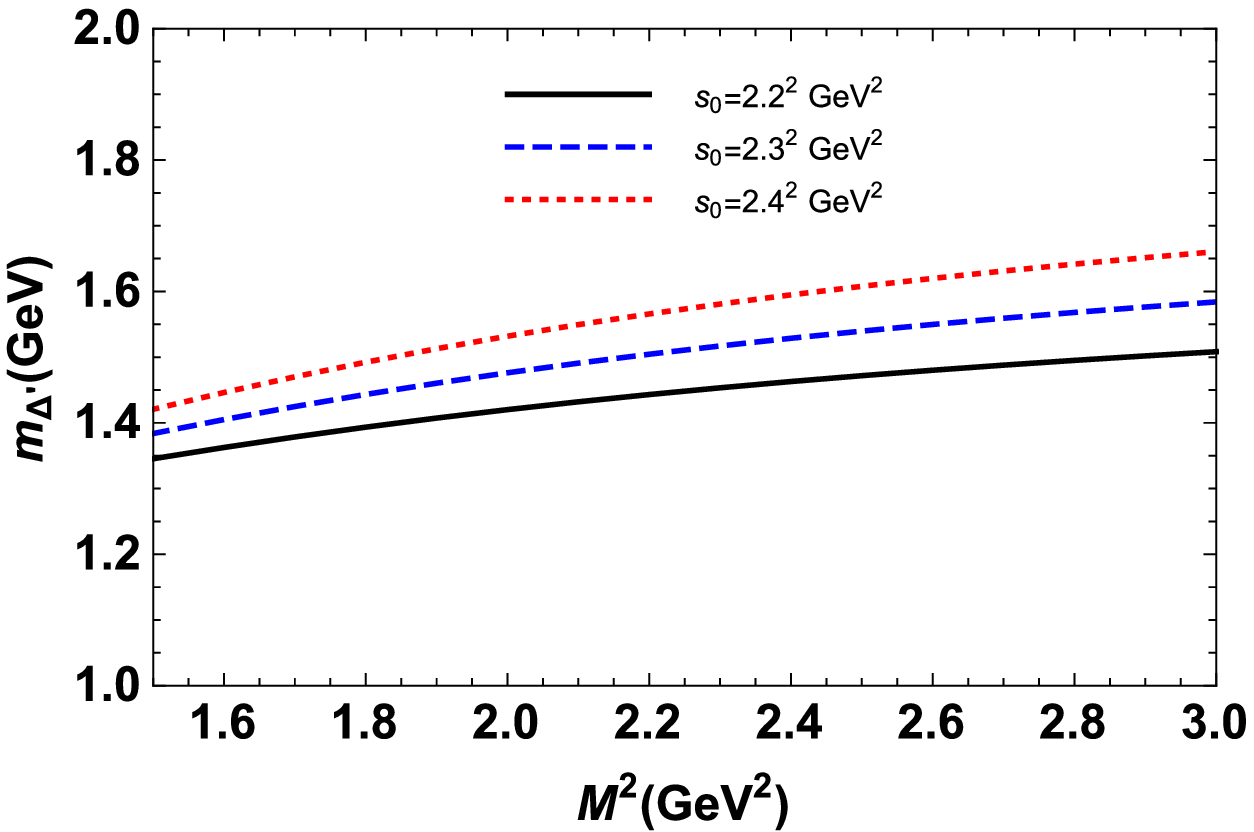}
\includegraphics[totalheight=5cm,width=7cm]{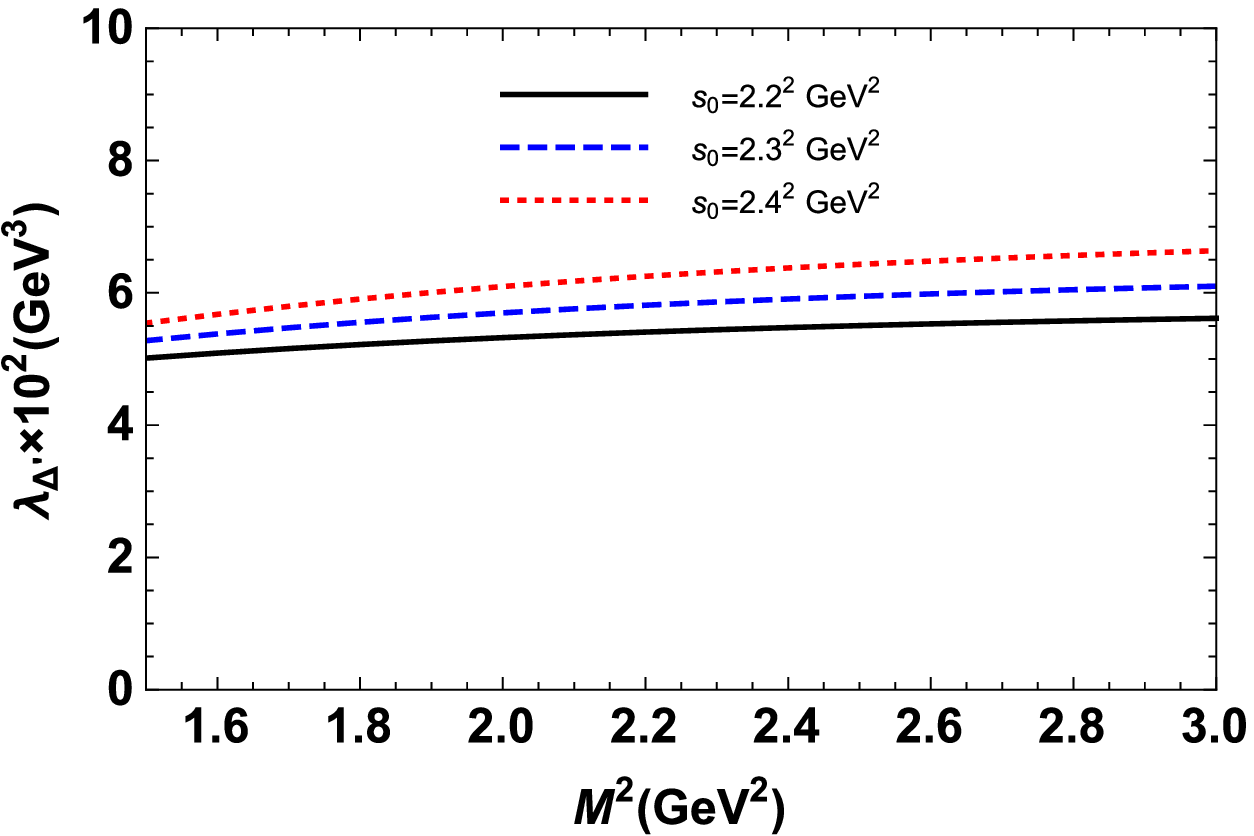}
\end{center}
\caption{\textbf{Left:} The mass of the radially excited $\Delta'$  baryon vs Borel
parameter $M^2$. \textbf{Right:}
 The residue  of the radially excited $\Delta'$  baryon vs Borel
parameter $M^2$. } \label{REDelta}
\end{figure}

\begin{figure}[h!]
\begin{center}
\includegraphics[totalheight=5cm,width=7cm]{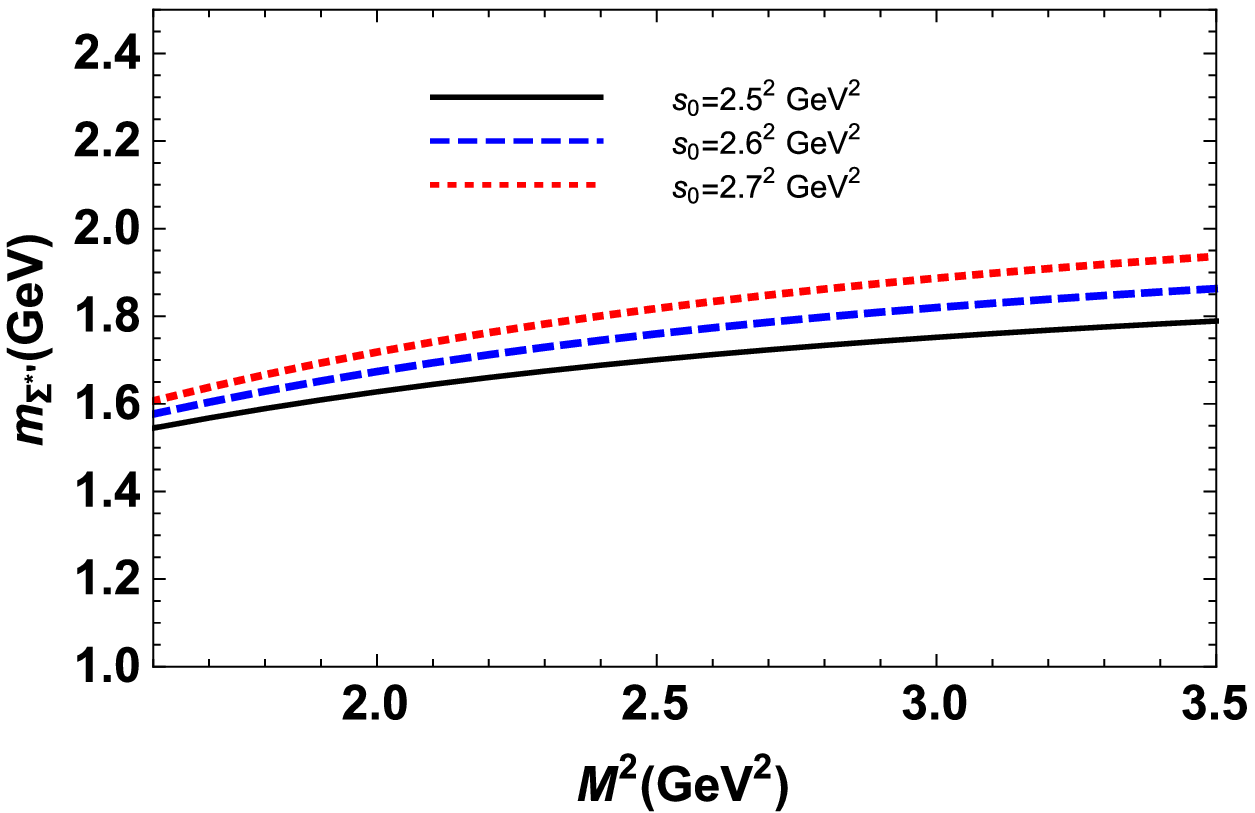}
\includegraphics[totalheight=5cm,width=7cm]{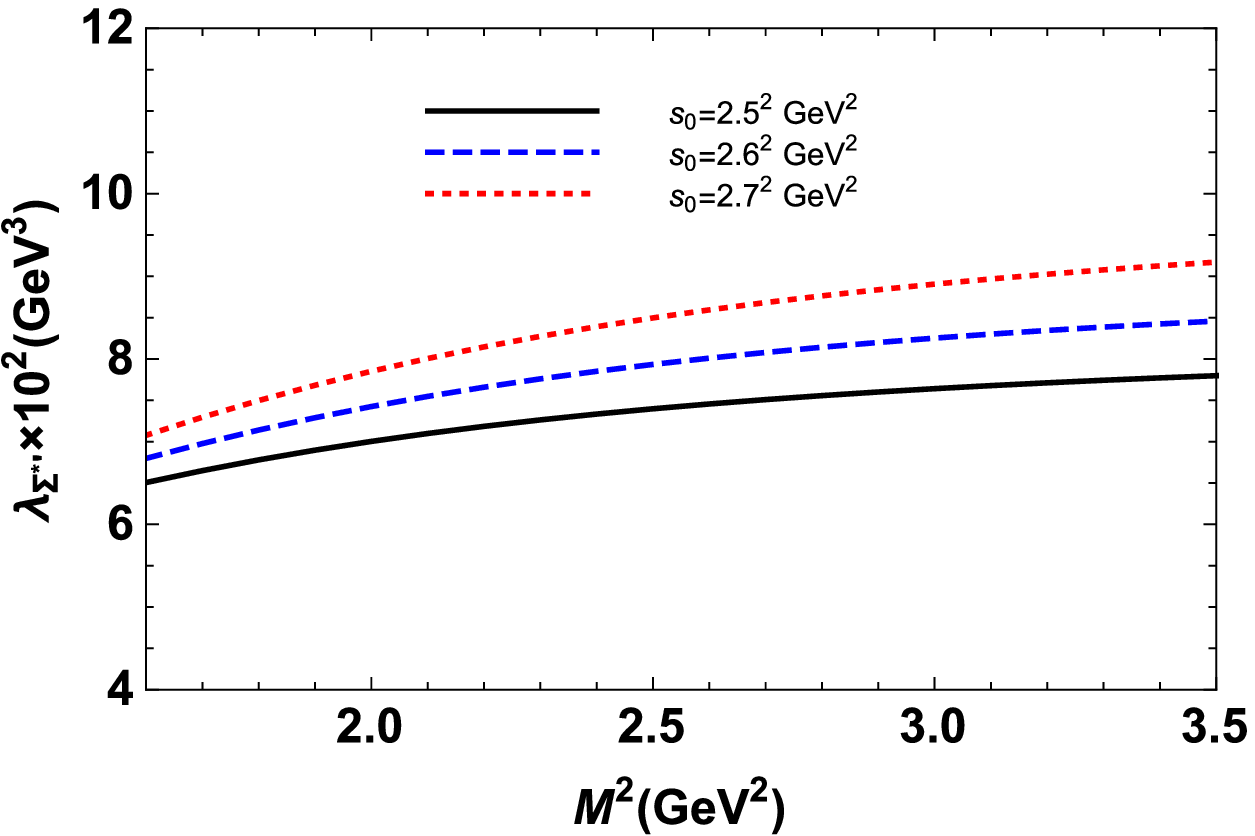}
\end{center}
\caption{The same as Fig. \ref{REDelta}, but for the radially excited $ \Sigma^{*'} $ baryon. } \label{RESigma}
\end{figure}

\begin{figure}[h!]
\begin{center}
\includegraphics[totalheight=5cm,width=7cm]{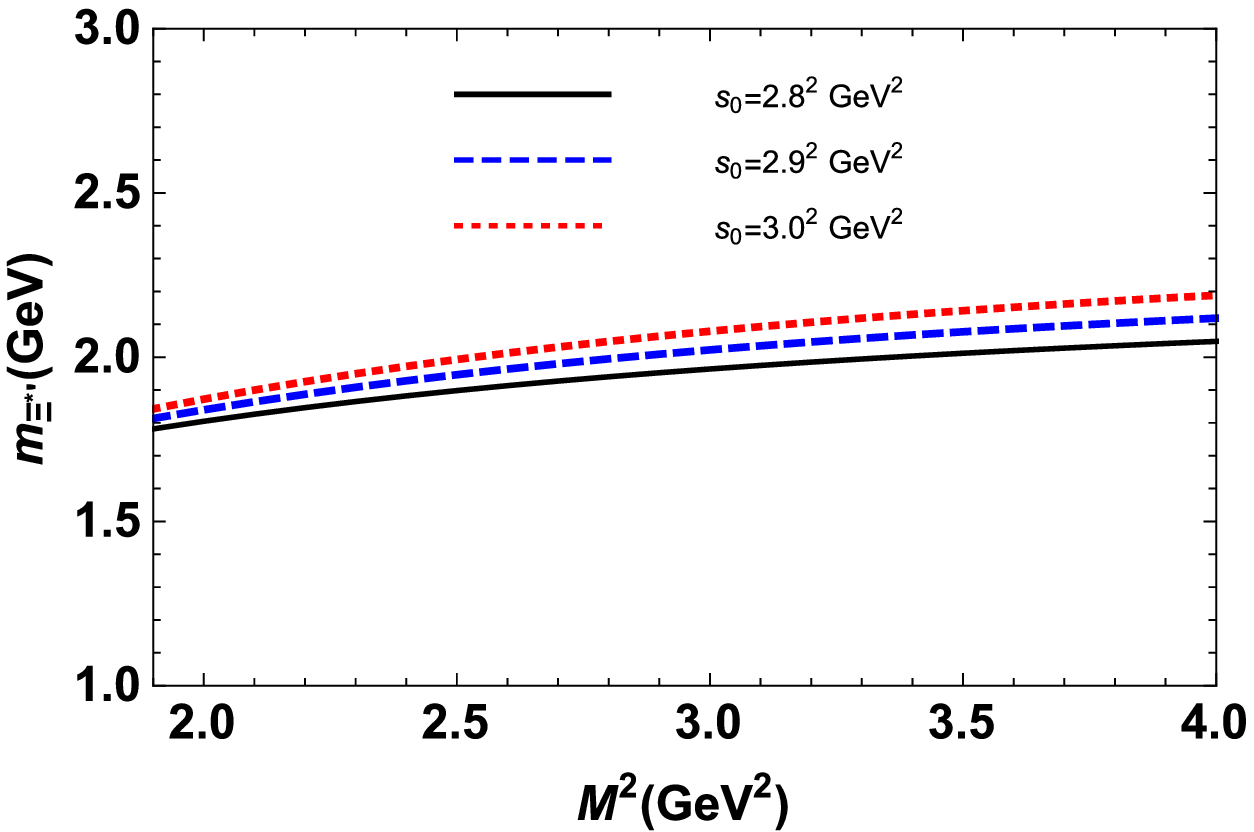}
\includegraphics[totalheight=5cm,width=7cm]{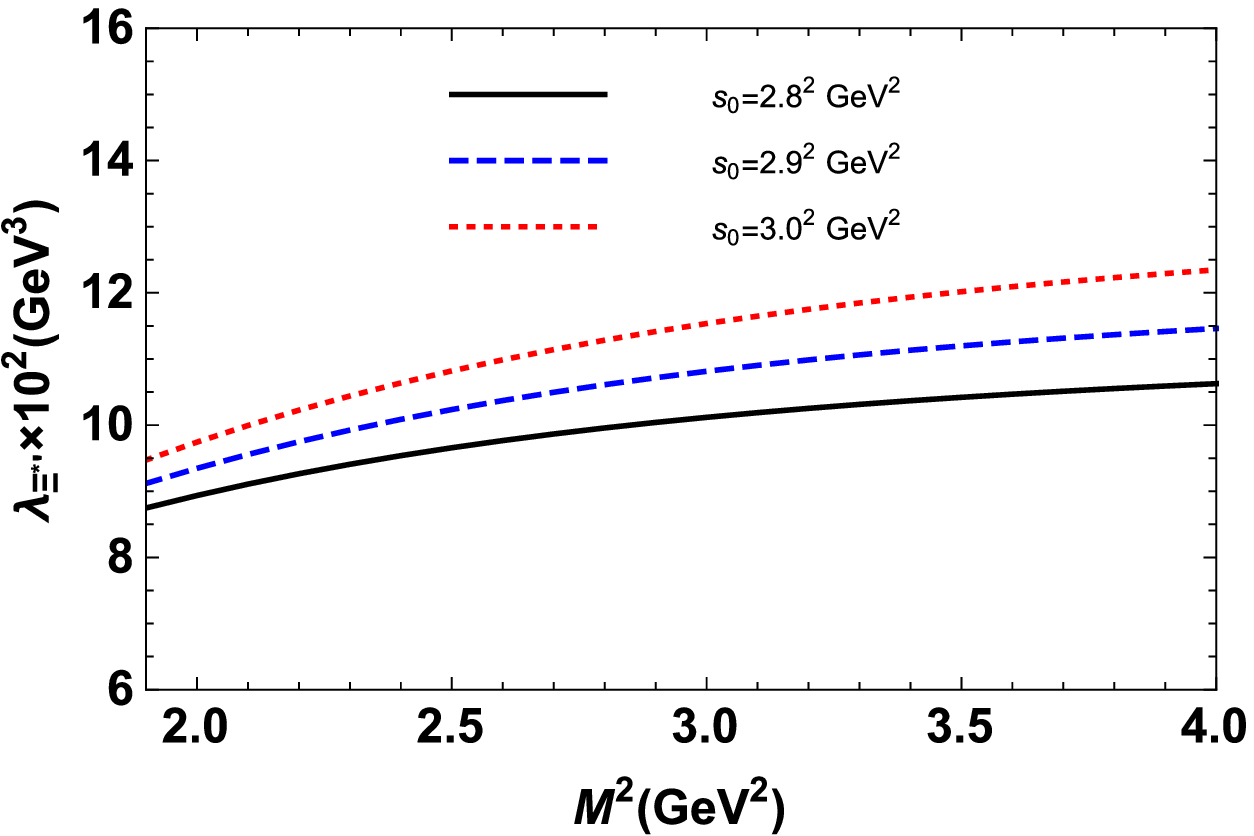}
\end{center}
\caption{The same as Fig. \ref{REDelta}, but for the radially excited  $ \Xi^{*'} $ baryon. } \label{REXi}
\end{figure}

\begin{figure}[h!]
\begin{center}
\includegraphics[totalheight=5cm,width=7cm]{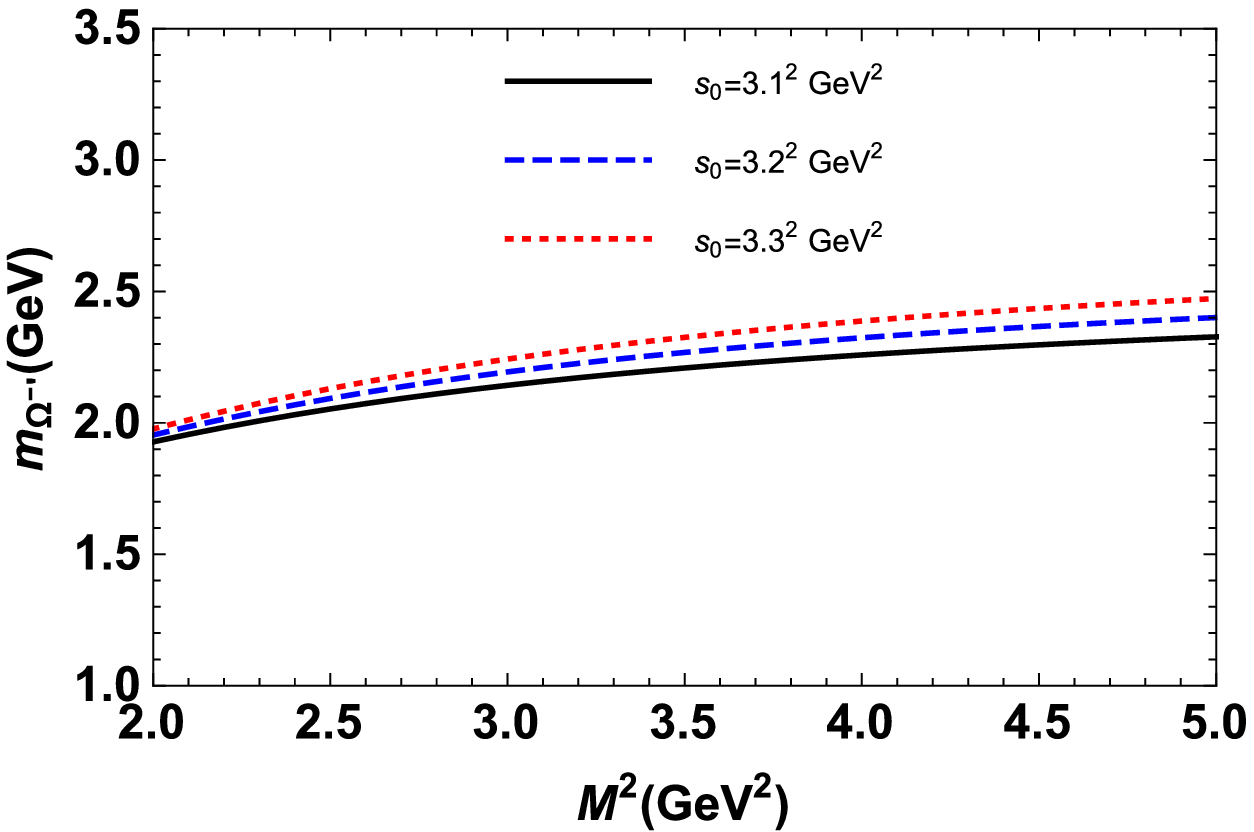}
\includegraphics[totalheight=5cm,width=7cm]{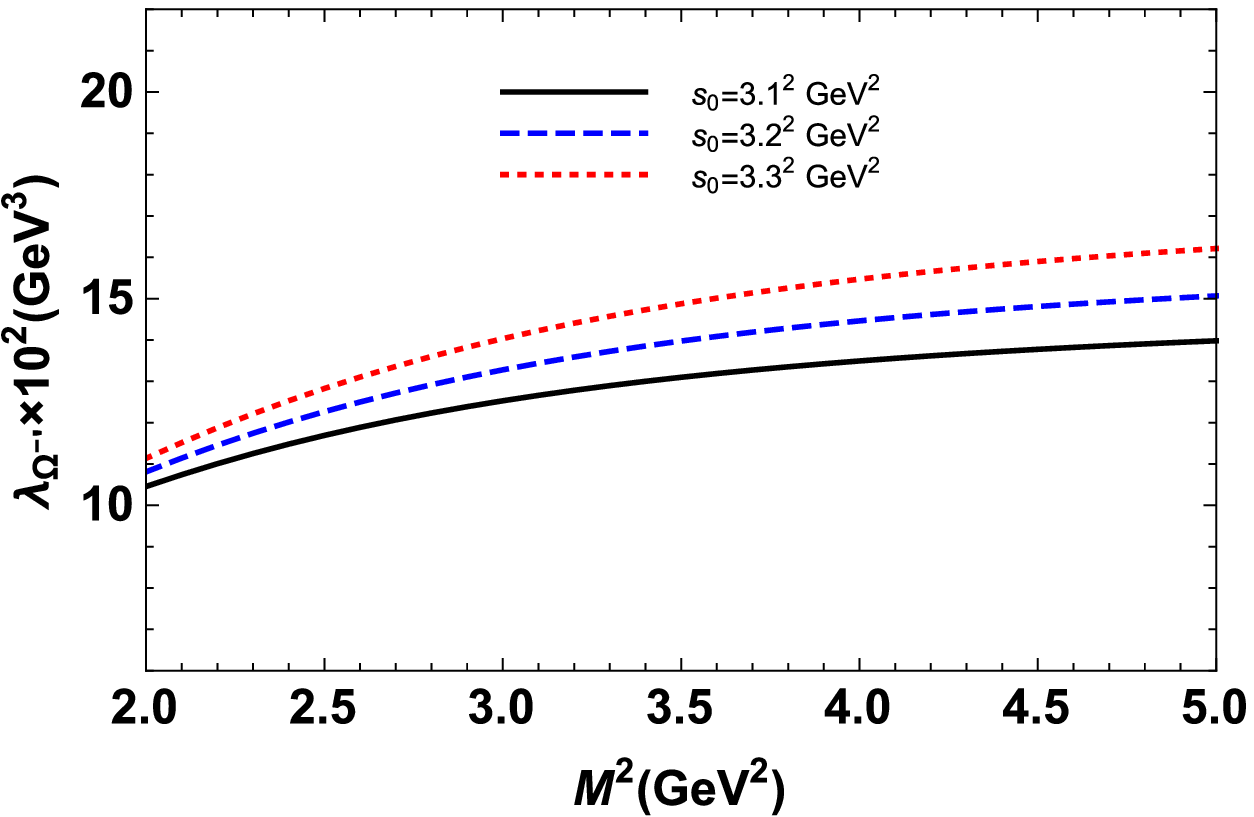}
\end{center}
\caption{The same as Fig. \ref{REDelta}, but for the radially excited  $ \Omega^{-'} $ baryon. } \label{REOmega}
\end{figure}

\end{widetext}

A quick glance at table \ref{tab:RDNumValues} leads to the following results:
\begin{itemize}
\item The values of the masses obtained for the radially excited $ \Delta'$ and $ \Sigma^{*'} $ baryons are in a good consistency with the experimentally well known $ \Delta(1600) $ and $ \Sigma(1730) $ baryons masses. 
\item Our result on the mass of the radially excited $ \Xi^{*'} $ baryon is in nice consistency with the experimentally observed $ \Xi(1950) $ state presented in PDG \cite{PDG}. The quantum numbers of this state is not established yet from the experiment. This consistency suggests that the $ \Xi(1950) $ state can be assigned as the first excited state of the ground state $ \Xi(1530) $. For more information about $ \Xi(1950) $ state see for instance \cite{Alitti:1968zz,Goldwasser:1970fk,Badier:1972az,DiBianca:1975ey,Briefel:1977bp,Biagi:1986vs,Adamovich:1999ic} and references therein.
\item In $ \Omega^{-} $ channel, our prediction for the excited state is also consistent with the mass of the experimentally observed (but unkwown quantum numbers) $ \Omega(2250)^{-} $ state within the errors. Therefore our analyses show that the $ \Omega(2250)^{-} $ listed in the PDG \cite{PDG} at $ \Omega^{-} $ channel can be assigned as the first excited state of the ground state $ \Omega^{-} $ with quantum numbers $ J^{P}=\frac{3}{2}^{+} $. For more information about  $ \Omega(2250)^{-} $ state see for instance \cite{ Biagi:1985rn,Aston:1987bb} and references therein.
\end{itemize}
%

\section{Conclusion}

\label{sec:Conc}

We have studied the spin--3/2,  $ \Delta $, $ \Sigma^{*} $, $ \Xi^{*} $ and $ \Omega^{-} $ baryons and calculated  the mass and residue of the corresponding ground and  first excited states in the framework of two-point QCD sum rules. First, we extracted the mass and residue of the ground state baryons by choosing an appropriate threshold from the obtained sum rules. We, then, used those values as inputs to obtain the mass and residue of the first excited state in each channels. Our results for the ground states are in nice agreements with the experimental data and the existing theoretical predictions. In the case of excited states, our predictions for the mass of the excited $ \Delta' $ and $ \Sigma^{*'} $ states are in good consistency with the experimentally known $ \Delta(1600) $ and $ \Sigma(1730) $ states' masses. In the case of excited $ \Xi^{*'} $ and $ \Omega^{-'} $ baryons, our results suggest that the experimentally poorly known $ \Xi(1950) $ and $ \Omega(2250)^{-} $ states can be assigned as the first excited states in $ \Xi^{*} $ and $ \Omega^{-} $ channels with $ J^{P}=\frac{3}{2}^{+} $. Our results for the residues can be verified via different theoretical approaches.


\section*{ACKNOWLEDGEMENTS}

Work of K.~A. was  financed by  Do\v{g}u\c{s} University under the grant BAP 2015-16-D1-B04.





\begin{thebibliography}{99}


\bibitem{Agashe:2014kda} 
  K.~A.~Olive {\it et al.} [Particle Data Group Collaboration],
  ``Review of Particle Physics,''
  Chin.\ Phys.\ C {\bf 38}, 090001 (2014).
  
\bibitem{Shifman:1978bx} 
  M.~A.~Shifman, A.~I.~Vainshtein and V.~I.~Zakharov,
  ``QCD and Resonance Physics. Theoretical Foundations,''
  Nucl.\ Phys.\ B {\bf 147}, 385 (1979).
  
\bibitem{Ioffe:1981kw} 
  B.~L.~Ioffe,
  ``Calculation of Baryon Masses in Quantum Chromodynamics,''
  Nucl.\ Phys.\ B {\bf 188}, 317 (1981)
  Erratum: [Nucl.\ Phys.\ B {\bf 191}, 591 (1981)].
  
\bibitem{Gelhausen:2014jea} 
  P.~Gelhausen, A.~Khodjamirian, A.~A.~Pivovarov and D.~Rosenthal,
  ``Radial excitations of heavy-light mesons from QCD sum rules,''
  Eur.\ Phys.\ J.\ C {\bf 74}, no. 8, 2979 (2014)
  [arXiv:1404.5891 [hep-ph]].
  
\bibitem{Jiang:2015paa} 
  J.~F.~Jiang and S.~L.~Zhu,
  ``Radial excitations of mesons and nucleons from QCD sum rules,''
  Phys.\ Rev.\ D {\bf 92}, no. 7, 074002 (2015)
  [arXiv:1508.00677 [hep-ph]].
  
\bibitem{Aliev}
  T.~M.~Aliev and S.~Bilmis,
  ``Analysis of radial excitations of octet baryons in QCD sum rules,''
  Adv.\ High Energy Phys.\  {\bf 2017}, 1350140 (2017)
  [arXiv:1612.09345 [hep-ph]].
  

\bibitem{Edwards:2012fx} 
  R.~G.~Edwards {\it et al.} [Hadron Spectrum Collaboration],
  ``Flavor structure of the excited baryon spectra from lattice QCD,''
  Phys.\ Rev.\ D {\bf 87}, no. 5, 054506 (2013)
  [arXiv:1212.5236 [hep-ph]].
  
\bibitem{Burch:2006cc} 
  T.~Burch, C.~Gattringer, L.~Y.~Glozman, C.~Hagen, D.~Hierl, C.~B.~Lang and A.~Schafer,
  ``Excited hadrons on the lattice: Baryons,''
  Phys.\ Rev.\ D {\bf 74}, 014504 (2006)
  [hep-lat/0604019].
  
\bibitem{Aznauryan:2009da} 
  I.~Aznauryan {\it et al.},
  ``Theory Support for the Excited Baryon Program at the Jlab 12- GeV Upgrade,''
  arXiv:0907.1901 [nucl-th].
  
\bibitem{Azizi:2016ddw}
  K.~Azizi and G.~Bozk{\i}r,
  ``Decuplet baryons in a hot medium,''
  Eur.\ Phys.\ J.\ C {\bf 76} (2016) no.10,  521
  [arXiv:1606.05452 [hep-ph]].
  

\bibitem{Belyaev:1982sa} 
  V.~M.~Belyaev and B.~L.~Ioffe,
  ``Determination of Baryon and Baryonic Resonance Masses from QCD Sum Rules. 1. Nonstrange Baryons,''
  Sov.\ Phys.\ JETP {\bf 56}, 493 (1982)
  [Zh.\ Eksp.\ Teor.\ Fiz.\  {\bf 83}, 876 (1982)].
  
  
\bibitem{Belyaev:1982cd} 
  V.~M.~Belyaev and B.~L.~Ioffe,
  ``Determination of the baryon mass and baryon resonances from the quantum-chromodynamics sum rule. Strange baryons,''
  Sov.\ Phys.\ JETP {\bf 57}, 716 (1983)
  [Zh.\ Eksp.\ Teor.\ Fiz.\  {\bf 84}, 1236 (1983)].
 
\bibitem{Wang:2010it} 
  Z.~G.~Wang,
  ``Analysis of the ${1/2^-}$ and ${3/2^-}$ heavy and doubly heavy baryon states with QCD sum rules,''
  Eur.\ Phys.\ J.\ A {\bf 47}, 81 (2011)
  [arXiv:1003.2838 [hep-ph]].
  
\bibitem{Wang:2010vn} 
  Z.~G.~Wang,
  ``Analysis of the ${3\over 2}^+$ heavy and doubly heavy baryon states with QCD sum rules,''
  Eur.\ Phys.\ J.\ C {\bf 68}, 459 (2010)
  [arXiv:1002.2471 [hep-ph]].
  
 
\bibitem{PDG} C. Patrignani et al. (Particle Data Group), "Review of Particle Physics", Chin. Phys. C, 40, 100001 (2016). 

\bibitem{Chetyrkin:2007vm}
  K.~G.~Chetyrkin, A.~Khodjamirian and A.~A.~Pivovarov,
  Phys.\ Lett.\ B {\bf 661} (2008) 250
  [arXiv:0712.2999 [hep-ph]].
  
\bibitem{Lee:1997ix}
  F.~X.~Lee,
  ``Predicative ability of QCD sum rules for decuplet baryons,''
  Phys.\ Rev.\ C {\bf 57} (1998) 322
  [hep-ph/9707332].
  
                   
  
\bibitem{Alitti:1968zz} 
  J.~Alitti {\it et al.},
  ``Evidence for Xi* Resonance with Mass 1930 MeV,''
  Phys.\ Rev.\ Lett.\  {\bf 21}, 1119 (1968).
  
\bibitem{Goldwasser:1970fk} 
  E.~L.~Goldwasser and P.~F.~Schultz,
  ``Xi- production in 5.5-gev/c k- p interactions,''
  Phys.\ Rev.\ D {\bf 1}, 1960 (1970).
  
\bibitem{Badier:1972az} 
  J.~Badier, E.~Barrelet, G.~R.~Charlton and I.~Videau,
  ``A search for xi* resonances in k- p interactions at 3.95 gev/c,''
  Nucl.\ Phys.\ B {\bf 37}, 429 (1972).
  
\bibitem{DiBianca:1975ey} 
  F.~A.~DiBianca and R.~J.~Endorf,
  ``Study of xi- and omega- Production from K- n and K- p Interactions at 4.93-GeV/c,''
  Nucl.\ Phys.\ B {\bf 98}, 137 (1975).
  
\bibitem{Briefel:1977bp} 
  E.~Briefel {\it et al.},
  ``Search for xi* Production in K- p Interactions at 2.87-GeV/c,''
  Phys.\ Rev.\ D {\bf 16}, 2706 (1977).
  
\bibitem{Biagi:1986vs} 
  S.~F.~Biagi {\it et al.},
  ``Xi* Resonances in Xi- Be Interactions. 2. Properties of Xi (1820) and Xi (1960) in the Lambda anti-K0 and Sigma-0 anti-K0 Channels,''
  Z.\ Phys.\ C {\bf 34}, 175 (1987).
\bibitem{Adamovich:1999ic} 
  M.~I.~Adamovich {\it et al.} [WA89 Collaboration],
  ``Production of Xi* resonances in Sigma- induced reactions at 345-GeV/c,''
  Eur.\ Phys.\ J.\ C {\bf 11}, 271 (1999)
  [hep-ex/9907021].
  
 
  
\bibitem{Biagi:1985rn} 
  S.~F.~Biagi {\it et al.},
  ``First Observation of $\Omega^*$ Resonances,''
  Z.\ Phys.\ C {\bf 31}, 33 (1986).
  
\bibitem{Aston:1987bb} 
  D.~Aston {\it et al.},
  ``Observation of $\Omega^*$- Production in $K^- p$ Interactions at 11-{GeV}/$c$,''
  Phys.\ Lett.\ B {\bf 194}, 579 (1987).

\end{thebibliography}
\end{document}